\documentclass[twocolumn]{aastex62}
\submitjournal{ApJL}
\newcommand  \oii   {\ifmmode \left[{\rm O}\,{\textsc ii}\right] \else [O\,{\sc ii}]\fi}

\newcommand  \civ  {\ifmmode {\rm C}\, {\sc IV}\ \else C\,{\sc IV}\fi}
\newcommand \hb    {\ifmmode {\rm H}\beta \else H$\beta$\fi}
\newcommand \ha    {\ifmmode {\rm H}\alpha \else H$\alpha$\fi}
\newcommand \hr    {\ifmmode {\rm H}\gamma \else H$\gamma$\fi}
\newcommand \hei    {\ifmmode {\rm He} {textsc{i}} \else He\,{\sc i}\fi}
\newcommand  \mgii  {\ifmmode {\rm Mg}{\textsc{ii}} \else Mg\,{\sc ii}\fi}
\newcommand{\ergs}{${\rm erg \ cm^{-2} \ s^{-1}}$ }
\newcommand{\erg}{${\rm erg \ s^{-1}}$ }

\newcommand{\kms}{\ifmmode {\rm km\,s}^{-1} \else km\,s$^{-1}$ \fi}
\newcommand  \oiii  {\ifmmode \left[{\rm O}\,{\textsc iii}\right] \else [O\,{\sc iii}]\fi}
\newcommand\Msun{$\rm M_{\odot}$}
\newcommand{\ergcmsA}{\ifmmode{\rm ergs}\, {\rm cm}^{-2}\,{\rm s}^{-1}\,{\rm\AA}^{-1} \else ergs\, cm$^{-2}$\, s$^{-1}$\, \AA$^{-1}$\fi}
\usepackage{CJK}
\usepackage{xcolor}
\usepackage{hyperref}
\hypersetup{
   colorlinks,
   linkcolor={blue!88!black!80},
   citecolor={blue!88!black!80},
   urlcolor={blue!88!black!80}
}
\shorttitle{\mgii\ CL AGNs and CL Unification}
\shortauthors{Guo et al.}
\begin{document}
\begin{CJK}{UTF8}{gbsn}

\title{Discovery of A \mgii\ Changing-look AGN and Its Implications for A Unification Sequence of Changing-look AGNs}

\author[0000-0001-8416-7059]{Hengxiao Guo (郭恒潇)}

\affiliation{Department of Astronomy, University of Illinois at Urbana-Champaign, Urbana, IL 61801, USA}
\affiliation{National Center for Supercomputing Applications, University of Illinois at Urbana-Champaign, Urbana, IL 61801, USA}

\author[0000-0002-0771-2153]{Mouyuan Sun}
\affiliation{CAS Key Laboratory for Researches in Galaxies and Cosmology, University of Sciences and Technology of China, Hefei, Anhui 230026, China}
\affiliation{School of Astronomy and Space Science, University of Science and Technology of China, Hefei 230026, China}
\affiliation{Department of Astronomy, Xiamen University, Xiamen, Fujian 361005, China}

\author[0000-0003-0049-5210]{Xin Liu}
\affiliation{Department of Astronomy, University of Illinois at Urbana-Champaign, Urbana, IL 61801, USA}
\affiliation{National Center for Supercomputing Applications, University of Illinois at Urbana-Champaign, Urbana, IL 61801, USA}

\author[0000-0002-1517-6792]{Tinggui Wang}
\affiliation{CAS Key Laboratory for Researches in Galaxies and Cosmology, University of Sciences and Technology of China, Hefei, Anhui 230026, China}
\affiliation{School of Astronomy and Space Science, University of Science and Technology of China, Hefei 230026, China}

\author{Minzhi Kong}
\affiliation{Department of Physics, Hebei Normal University, No. 20 East of South 2nd Ring Road, Shijiazhuang 050024, China}
\affiliation{Department of Astronomy, University of Illinois at Urbana-Champaign, Urbana, IL 61801, USA}

\author{Shu Wang}
\affiliation{Kavli Institute for Astronomy and Astrophysics, Peking University, Beijing 100871, China}
\affiliation{Department of Astronomy, School of Physics, Peking University, Beijing 100871, China}

\author[0000-0001-6938-8670]{Zhenfeng Sheng}
\affiliation{CAS Key Laboratory for Researches in Galaxies and Cosmology, University of Sciences and Technology of China, Hefei, Anhui 230026, China}
\affiliation{School of Astronomy and Space Science, University of Science and Technology of China, Hefei 230026, China}

\author{Zhicheng He}
\affiliation{CAS Key Laboratory for Researches in Galaxies and Cosmology, University of Sciences and Technology of China, Hefei, Anhui 230026, China}
\affiliation{School of Astronomy and Space Science, University of Science and Technology of China, Hefei 230026, China}

\email{hengxiao@illinois.edu (H.X.G.), ericsun@ustc.edu.cn (M.Y.S.),\\ confucious\_76@163.com (M.Z.K.)}

\begin{abstract}
Changing-Look (CL) is a rare phenomenon of Active Galactic Nuclei (AGNs), which exhibit emerging or disappearing broad lines accompanied by continuum variations on astrophysically short timescales ($\lesssim$ 1 yr to a few decades). While previous studies have found Balmer-line (broad H$\alpha$ and/or H$\beta$) CL AGNs, the broad \mgii\ line is persistent even in dim states. No unambiguous \mgii\ CL AGN has been reported to date. We perform a systematic search of \mgii\ CL AGNs using multi-epoch spectra of a special population of \mgii-emitters (characterized by strong broad \mgii\ emission with little evidence for AGN from other normal indicators such as broad H$\alpha$ and H$\beta$ or blue power-law continua) from the Fourteenth Data Release of the Sloan Digital Sky Survey. We present the discovery of the first unambiguous case of a \mgii\ CL AGN, SDSS J152533.60+292012.1 (at redshift $z$ = 0.449), which is turning off within rest-frame 286 days. The dramatic diminishment of \mgii\ equivalent width (from 110 $\pm$ 26 \AA\ to being consistent with zero), together with little optical continuum variation ($\Delta V_{\rm max-min}$ $=$ 0.17 $\pm$ 0.05 mag) coevally over $\sim$ 10 years, rules out dust extinction or a tidal disruption event. Combined with previously known \hb\ CL AGNs, we construct a sequence that represents different temporal stages of CL AGNs. This CL sequence is best explained by the photoionization model of \cite{Guo19}. In addition, we present two candidate turn-on \mgii\ CL AGNs and a sample of 361 \mgii-emitters for future \mgii\ CL AGN searches.
\end{abstract}


\section{Introduction} 
\label{sec:intro}
%describe the background of CL AGNs 
CL is a useful phenomenon to understand the physical structure of AGNs and a natural laboratory to explore the evolution between AGNs and normal galaxies. Despite the massive modern spectroscopic/photometric sky surveys, only dozens of Balmer-line CL AGNs have been discovered with type transition timescale ranging from several months to decades \citep{LaMassa15,Runco16,Runnoe16,Ruan16b}, leading to a detection rate much smaller than 1\% \citep{Macleod16,Yang18}. The intrinsic nature of this rapid CL behavior was usually explained by dust reddening \citep[e.g.,][]{Goodrich89,Tran92}, accretion rate change \citep[e.g.,][]{LaMassa15}, or Tidal Disrupted Event (TDE) \citep{Merloni15}. However, recent evidence (e.g., the polarization observation in \cite{Hutsemekers17} and mid-infrared echo in \cite{Sheng17}) suggests that the variation of accretion rate is likely to be the primary origin for CL AGNs, although the short transition timescale challenges the standard thin disk model \citep{Shakura73} which predicts a transition timescale of $\sim10^{4}$ yrs \citep{Macleod16}. In order to address the timescale problem, competing models, e.g., magnetically elevated disk model \citep{Dexter19b}, and instabilities arising from magnetic torque near the event horizon \citep{Ross18} are proposed. On the other hand, repeating X-ray observations suggest that the CL phenomenon in supermassive black hole might be analogous to the structure of accretion flows in stellar-mass black holes \citep{Ruan19}. 

%describe the particularity of mgii line
Previous observations have revealed the emerging or disappearing of broad Balmer lines (e.g., \hb\ or \ha) in CL AGNs, whereas the broad \mgii\ is always persistent even in the dim state \citep{Macleod16,Macleod19,Yang18,Yang19}. To date, no unambiguous \mgii\ CL phenomenon has been reported yet. 

On the other hand, \cite{Roig14} discovered $\sim$ 300 unusual broad \mgii-emitters. These sources show strong and broad \mgii\ line, but very weak emission in other normal indicators of AGN activity, like \ha, \hb, and near-ultraviolet power-law continuum. They considered these \mgii-emitters as a potentially new class of AGNs. However, we argued that they are more likely to be the transition stage in CL AGNs \citep[][also see \S \ref{sec:cl_seq}]{Guo19}.

The difficulty of discovering \mgii\ CL might mainly be caused by the weak variability of \mgii\ line \citep{Macleod19,Yang19}. Previous reverberation mapping programs encountered a similar situation that the response of \mgii\ line to continuum variation is often undetectable \citep[e.g.,][]{Cackett15} except for a few sources \citep[e.g.,][]{Clavel91}. Two possible mechanisms are proposed to explain the weak variability and the lack of response to continuum fluctuations: 1) geometric dilution that makes the relative outer \mgii\ emitting region to get only some of the scatter continuum emission \citep{Sun15}, or 2) the intrinsic slow response of \mgii\ dominated by atomic physics and radiative transfer within the line-emitting clouds \citep{Goad93,Korista00,Guo19}.

% in this work...
In order to understand the phenomena of CL AGNs and \mgii-emitters, as well as the radiative mechanism of \mgii\ line, which is an important proxy of the black hole mass at quasar activity peak (i.e., 1 $<$ z $<$2 ), we performed a series of work. In \cite{Guo19}, we first quantitatively compared the line-variability behaviors between \mgii\ and Balmer lines and demonstrated a good consistency of CL phenomenon with the photoionization models. In this letter, we present the results of the first systematic search of \mgii\ CL AGNs by studying repeat spectra of \mgii-emitters from SDSS DR14. In particular, we present the discovery of the first unambiguous case of \mgii\ CL. 

The paper is organized as follows. In \S \ref{sec:data}, we describe the data and sample selections for \mgii-emitters. In \S \ref{sec:results}, we present an unambiguous \mgii\ CL, as well as two candidates of \mgii\ CL AGN. Then we construct the observed CL sequence and discuss its implications. Finally, we draw our conclusion and discuss the future work in \S \ref{sec:con}. Throughout this paper, a cosmology with $H_{0}$ = 70 $\rm km\,s^{-1}Mpc^{-1}$, $\Omega_{m} = 0.3$ and $\Omega_{\Lambda} = 0.7$ was adopted.

\section{Data and Sample Selection} \label{sec:data}
\subsection{SDSS spectrum}
All the spectra in this work are obtained from the public SDSS DR14 database \citep{Abolfathi18}, which covers 14,555 deg$^2$. Benefit from its $\sim$ 20-year cumulative data, extensive multi-epoch spectra are quite suitable to investigate the AGN spectral variability. The multi-epoch spectroscopic observations are mainly from three parts: 1) the overlapped survey areas between adjacent plates; 2) dedicated programs, e.g., Time Domain Spectroscopic Survey \citep{Ruan16a} and SDSS reverberation mapping \citep{Shen15}; 3) re-observed plates due to insufficient Signal-to-Noise Ratio (SNR). The spectral wavelength coverage for SDSS I\&II (SDSS III) is 3800 -- 9200 (3600 -- 10400) \AA\ with spectral resolution R $\sim$ 1850 -- 2200, and the five-band $ugriz$ magnitudes have typical errors of about 0.03 mag in depth to 22.0, 22.2, 22.2, 21.3, 20.5 mag \citep{Abazajian09}.

\subsection{CSS light curve}
Although the optical light curves are not used to select \mgii\ CL AGNs in \S \ref{sec:selection}, they are still useful for understanding the origins of the CL behavior. The Catalina Sky Survey \citep[CSS,][]{Drake09} repeatedly covered 26,000 deg$^2$ on the sky using a 0.7 m Schmidt telescope with a wide field of view of 8.1 deg$^2$. The photometric data were unfiltered and calibrated to V-band magnitude, to a depth of $\sim$ 20 mag.

%The Palomar Transient Factory \citep[PTF,][]{Law09} was an optical synoptic survey to explore the transient and variable sky. The observations were made at Palomar Observatory by the 1.2 m Samuel Oschin Schmidt telescope with the CHF12K camera, providing a wide field of view of 7.26 deg$^2$. It covered $\sim$3000 deg$^2$ of the sky with a 5$\sigma$ limiting magnitude of $\sim$20.6 in Mould-$R$ and $\sim$21.3 in SDSS-$g$ bands with an average 5 day cadence. 

\subsection{Sample selection of \mgii-emitters} \label{sec:selection}
Previous observations of CL AGNs indicate that the so-called \mgii-emitters are likely to be the faint states of \hb\ CL AGNs \citep{Macleod16,Yang18,Macleod19}. In order to search both \mgii\ and \hb\ CL AGNs, we define the \mgii-emitters as those with prominent broad \mgii\ but no broad \hb\ component ($FWHM_{\hb}$ $<$ 1000 \kms), similar to \cite{Roig14}. The advantage of this approach is able to discover both \mgii\ and \hb\ CL AGNs based on the multi-epoch spectra when AGNs turn off or on. Compared with the widely used variability-color selection \citep{Macleod16, Sheng17,Macleod19,Yang18}, our method 
servers as a tailored approach for searching \mgii\ CL AGNs at low luminosity end, since all the \mgii-emitters are very faint (see below).

\begin{figure*}
\centering
\hspace*{-1cm}
\includegraphics[width=20cm]{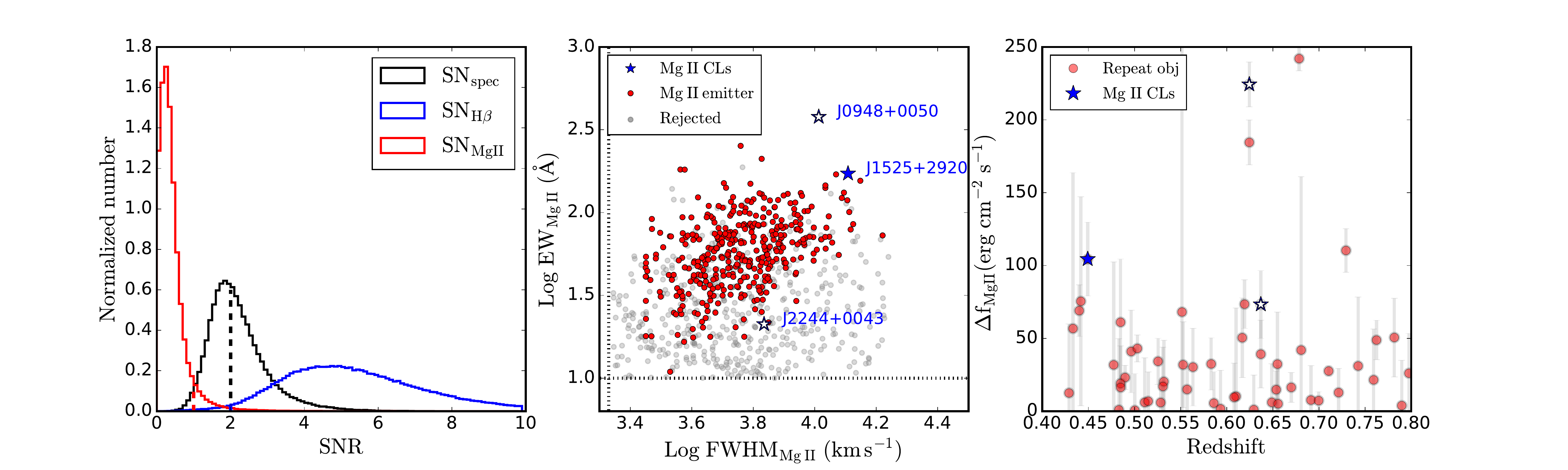}
\caption{Left panel: distributions of the SNR for \mgii\ (red), \hb\ (blue) and whole spectrum (black). About 16000 \mgii-emitter candidates with $SN_{\rm MgII}> $1 and both $ SN_{\rm spec}$ and $SN_{\rm H\beta} >$ 2 are selected from the SNR cut. Middle panel: The FWHM-EW diagram of \mgii\ line. We complied a sample of 361 \mgii-emitters (red dots) based on our criteria in \S \ref{sec:selection}. One unambiguous (filled blue star) and two tentative \mgii\ CL AGNs (empty blue stars) are marked. The grey dots are rejected candidates due to noise fitting or with significant broad \hb\ component. Two black dotted lines are the lower limits of EW ($>$ 10 \AA) and FWHM ($>$ 2000 \kms) to exclude the potential Type II AGNs and reduce the contamination from noise fitting. Right panel: \mgii\ line variability as a function of redshift. We find three \mgii\ CL AGNs out of ten sources with significant \mgii\ variability ($> 3\sigma$) in 52 \mgii-emitters with repeat observations.     
}
\label{fig:selection}
\end{figure*}

We start with all spectra (4.8 million) in SDSS DR14 database \citep{Abolfathi18}. Followings are the selection criteria for \mgii-emitter candidates: 
\begin{enumerate}
\item Redshift: 0.4 $<$ z $<$ 0.8, zWarning = 0
\item Class = ``QSO" or ``GALAXY"
\item Mg II flux $>$ 0, and ($FWHM_{\rm \hb}$ $<$ 1000 \kms or \hb\ flux $<$ 0)
\item $SN_{\rm spe} >2$, $SN_{\rm \hb} >2$ and $SN_{\rm \mgii} >1$ 
%\item 2000 \kms $<$ $FWHM_{\rm MgII}$ $<$ 20000 \kms\ and $EW_{\rm MgII}$ $>$ 10\AA.
\end{enumerate}
To include the \mgii\ line, \hb\ - \oiii\ complex of AGNs, and simultaneously avoid the emission lines to be too close to the spectral edges resulting in low SNRs, Criteria 1 \& 2 are applied. In addition, Criterion 3 ensures that the \mgii\ (\hb) is emission line (narrow emission line or absorption line) based on the measurements from SDSS automatic pipeline \citep{Bolton12}. As shown in left panel of Figure \ref{fig:selection}, the SNRs of continuum and lines (Criterion 4) are needed to ensure the spectral qualities of the candidates. We note that all these candidates are very faint with a typical flux density of $10^{-17}$ \ergcmsA\ at 3000\AA, thus the SNRs are lower than the typical values of ordinary quasars \citep[e.g., 5 $\sim$ 10 in ][]{Shen11}. These four criteria yield $\sim$ 16000 \mgii-emitter candidates. 

Then we use PyQSOFit\footnote{A public python code for quasar spectral fitting, see https://github.com/legolason/PyQSOFit.} \citep{Guo18,Shen19} to perform the local fit for the \mgii\ region ([2700, 2900]\AA) with a power-law continuum, Iron template and up to three Gaussian profiles to extract the line properties. To exclude the potential Type II AGNs with narrow \mgii\ doublets and also alleviate the noise fitting, we select the candidates with 
\begin{enumerate}
\item[5.] 2000 \kms $<$ $FWHM_{\rm MgII}$ $<$ 20000 \kms\ and $EW_{\rm MgII}$ $>$ 10\AA.
\end{enumerate}
This leaves $\sim$ 800 objects (red and grey dots), shown in middle panel of Figure \ref{fig:selection}.

Next, we visually inspect each spectrum to confirm that the spectral fitting for \mgii\ line and the measurements of $\hb$ line from SDSS automatic pipeline are reasonable. About half of $\sim$ 800 objects, usually located in the lower right portion of FWHM-EW diagram, were excluded because of the extremely week broad Mg II line blending with the continuum, which can significantly affect the spectral decomposition. Another $\sim$ 50 objects showing significant broad \hb\ components\footnote{For these sources, the SDSS automatic pipeline fitting results are biased; therefore, we refit their \hb-\oiii\ complex for further confirmation.} are also excluded, which are usually with large $EW_{\rm \mgii}$ located in the upper portion of the FWHM-EW diagram. This process leaves us a sample of 361 (with a detection rate of $\sim$0.02\% in $\sim$ 2 million galaxies/quasars) unique \mgii-emitters\footnote{The \mgii-emitter catalog is available here: https://github.com/legolason/MgII-emitter-catalog. \\Through privately obtained \mgii-emitter catalog from \cite{Roig14}, we found that the overlaps between two catalog are less than 10\% due to the different selection criteria.} including 52 objects with multi-epoch observations. Their spectra also usually show significant galactic features, e.g., strong absorption lines (e.g., Ca H+K), 4000 \AA\ break and weak power-law continuum.

Finally, we refit the brightest and faintest epochs of these 52 \mgii-emitters, and find 10 objects with significant \mgii\ variability ($>$ 3$\sigma$) in the right panel of Figure \ref{fig:selection}. Rejected seven ordinary sources due to normal broad \mgii\ variation without CL behavior, it leaves three \mgii\ CL AGNs (see Figure \ref{fig:selection}, \ref{fig:spectrum}, \ref{fig:2MgII_CL} \& Table \ref{tab:table1}), i.e., a detection rate of $\sim$0.001\% (3/52 $\times$ 0.02\% ) based on \mgii-emitters, which is consistent with that of \hb\ CL AGNs in \cite{Yang18}. We also discovered new \hb\ CL AGNs, which will be shown in a future paper.

\section{Results and Discussion}
\label{sec:results}
\subsection{Discovery of the first \mgii\ CL AGN}
\begin{figure*}[h]
\centering
\includegraphics[width=18cm]{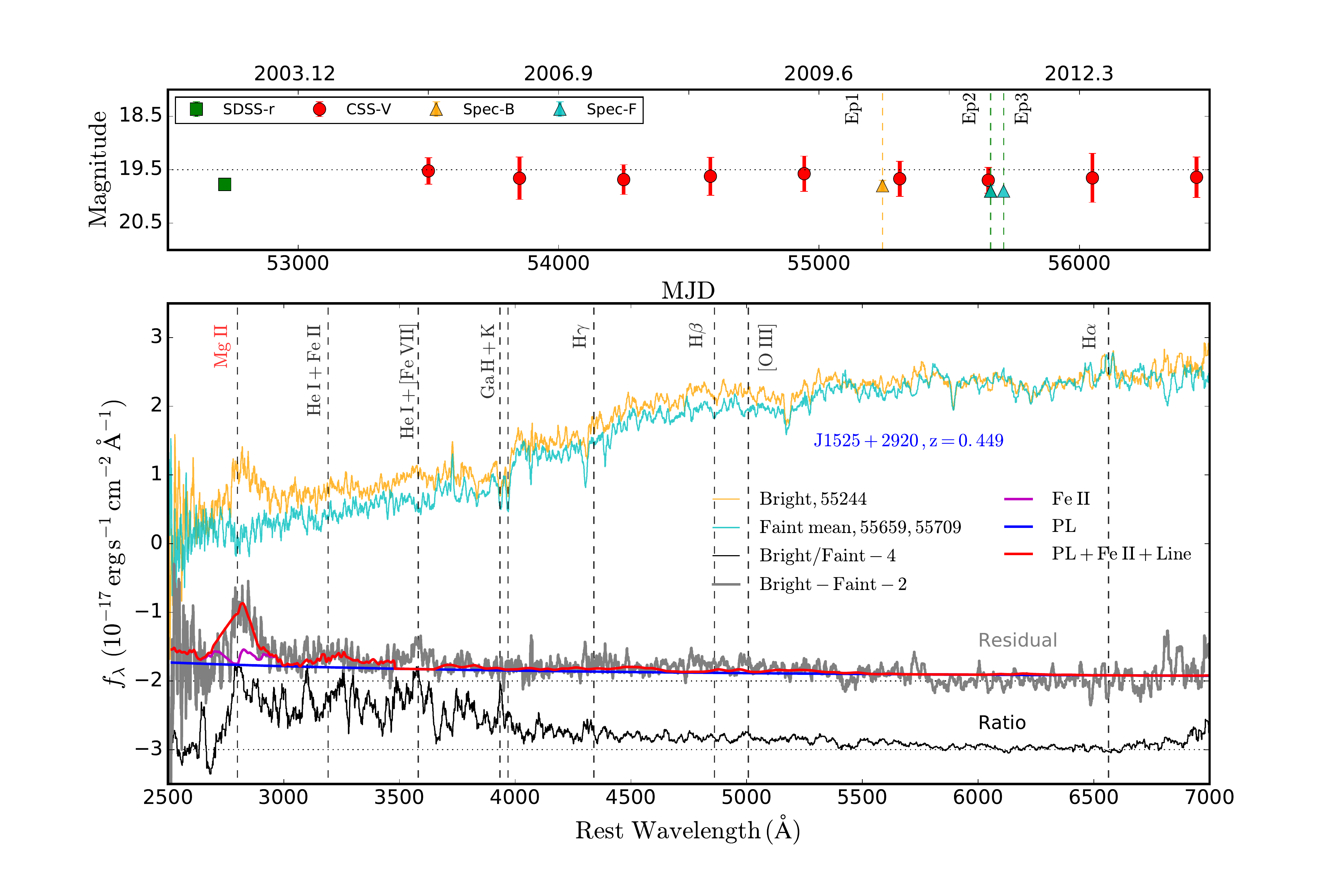}
\caption{The unambiguous turn-off CL phenomenon of J1525+2920. The bottom panel shows the bright state (orange) and faint state (green, the mean of two faint epochs to improve the SNRs) with box-car smoothing of 10 pixels for clarity. The broad \mgii\ line disappears accompanied by Helium and Iron lines (3191 \AA\ and 3581 \AA) with fading continuum in 286 days in rest frame. The residuals (grey) are well fitted by a model (red) consisting of a power-law (blue), Iron template (magenta), and one Gaussian profile. The spectral ratio (black) of the bright and faint states shows that the relatively larger variability in the blueward. The dotted lines under the spectral ratio and the residuals are added to guide the eye, and the dashed lines mark the prominent lines in UV/optical bands. The top panel presents the seasonally averaged CSS light curve, together with photometries from three SDSS spectra and SDSS-$r$. Only little variations (0.17 $\pm$ 0.05 mag) can be detected on a timescale of $\sim$ 10 yrs. The corresponding spectra are marked with dashed orange/green lines.}
\label{fig:spectrum}
\end{figure*}

\begin{figure*}
\centering
\hspace*{-1cm}
\includegraphics[width=15.cm]{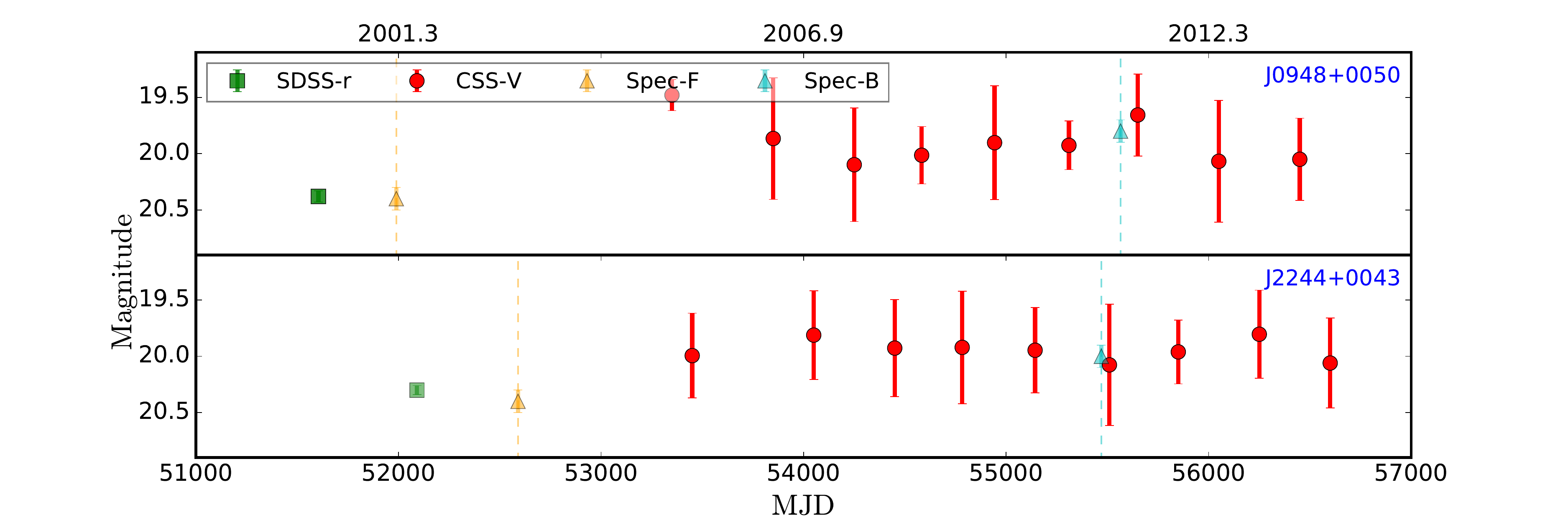}
\hspace*{-1cm}
\includegraphics[width=15.cm]{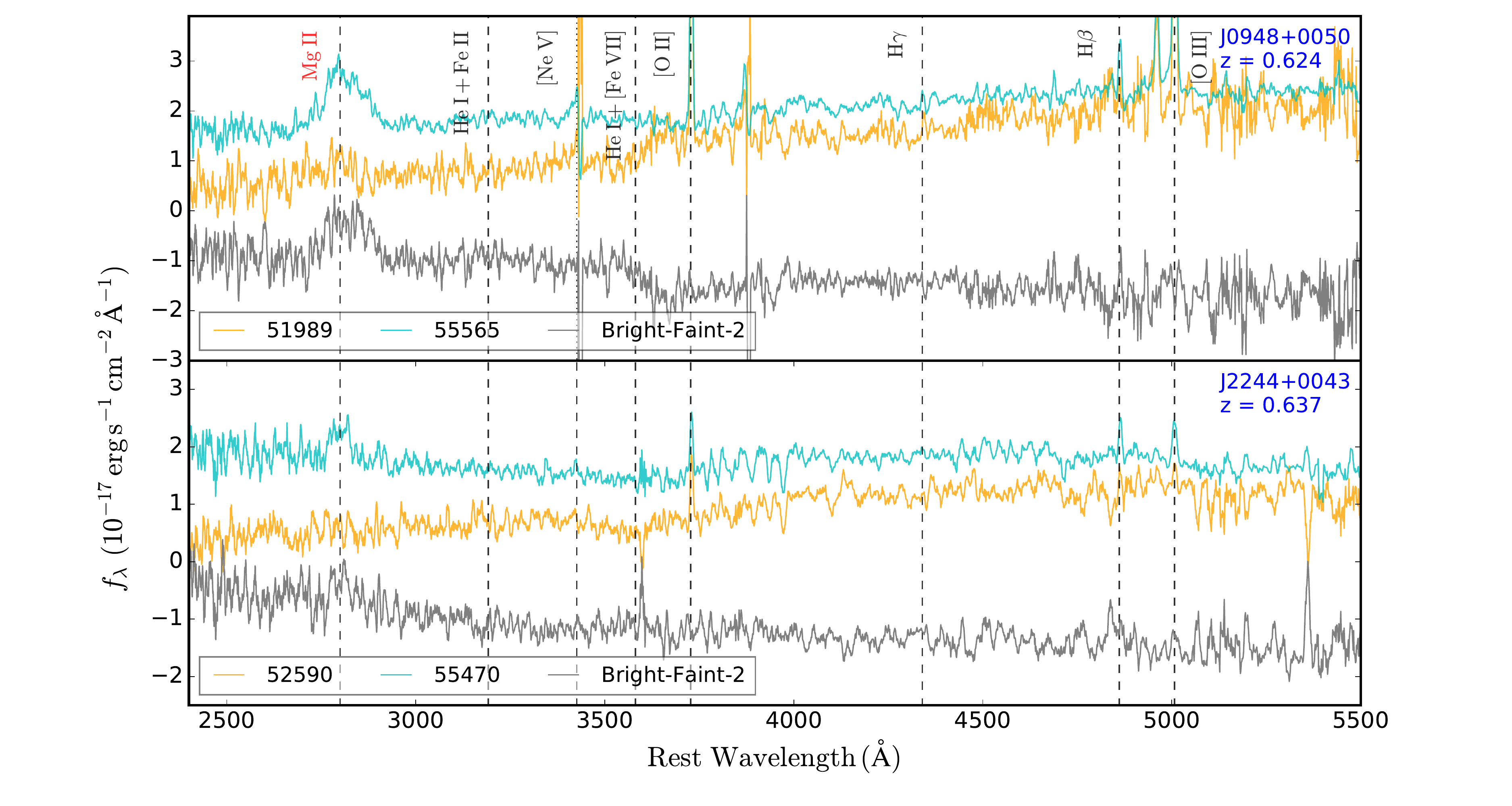}
\caption{{Two tentative turn-on \mgii\ CL AGNs.} 
Bottom: Both bright states (green) are selected as \mgii\ emitters. With continuum turns on, J0948+0050 presents a strong broad \mgii\ within 2202 rest-frame days, while J2244+0043 shows a weak broad \mgii\ and disappearance of the absorption features of Balmer lines within 1759 rest-frame days. For clarity all the spectra are smoothed with a box-car of 10 pixels. Upper: The multi-survey light curves show the maximum variabilities of $\sim$ 1 mag and $\sim$ 0.5 mag for J0948+0050 and J2244+0043, respectively. }
\label{fig:2MgII_CL}
\end{figure*}

Figure \ref{fig:spectrum} shows an unambiguous turn-off \mgii\ CL AGN (J1525+2920) at z = 0.449. This object was first selected as a \mgii-emitter in the bright state by our work, which was targeted as a luminous red galaxy by SDSS. J1525+2920 shows a dramatic change in \mgii\ equivalent width (EW), i.e., $EW_{\rm Mg\;II}$ = 110 $\pm$ 26 \AA\ to 0 (or $\Delta f_{\mgii}$ = 103 $\pm$ 25 \ergs\ at 4$\sigma$ level), with a factor of 2 continuum variation blueward of rest-frame 4000\AA, which rules out the dust reddening scenario for CL behavior as this model expects a constant line $EW_{\mgii}$. Moreover, the accompanied disappearing of Helium and Iron lines at 3191 \AA\ and 3581 \AA, as typical features of CL AGNs and TDEs \citep{Yan19,Brown16}, further supports that this is a genuine \mgii\ CL event rather than a false alarm due to the calibration problem\footnote{We checked the quality of the plates are good, and other objects in these plates are normal.}. The self-consistence of two faint epochs\footnote{See all spectra here: http://skyserver.sdss.org/ with RA, DEC = (15:25:33.60, +29:20:12.12)} also suggests that the SDSS flux calibration is robust for this object. The residual spectrum (bright $-$ faint) is well fitted by a power-law continuum $\rm f_{\lambda} = \lambda^{-1.24\pm0.05}$, which may indicate a possible AGN origin of the varying component\footnote{The typical optical slope of AGN is $\rm f_{\lambda} = \lambda^{-1.54\pm0.49}$, see \cite{Guo16} for details.}. Its rest-fame transition timescale is less than 286 days, which is consistent with other normal \hb\ CL AGNs \citep{Macleod16,Yang18}.

By convolving with the SDSS filters, this source shows the variations in $g-$ and $r-$band are $0.54\pm0.02$ mag and $0.1\pm 0.01$ mag  respectively, which would be missed by conventional variability selections \citep[e.g., $\Delta$g $>$ 1 mag,][]{Macleod16,Rumbaugh18}.   

The seasonally averaged CSS light curve in the upper panel of Figure \ref{fig:spectrum}, together with these $r-$band synthetic magnitudes obtained from three spectra and SDSS-$r$, indicates a weak variability ($\Delta V_{\rm max-min}$ = 0.17 $\pm$ 0.05 mag) over 10 yrs. This strongly disfavors the TDE scenario, which typically exhibits a rapid raising phase with several magnitudes and a slow decay by $\sim$ $t^{-5/3}$ within at most several years \citep{Rees88,Evans89}, as well as some temporal supernova-driven broad emission lines \citep{Simmonds16}. The slight difference in the photometric systems of SDSS and CSS can be safely ignored for our purposes.

Given the $FWHM_{\mgii} = 12700 \pm 300\ \kms$, we estimate the black hole mass of $M_{\rm BH} = 10^{8.0 \pm 0.1}$ \Msun\ (1 $\sigma$ statistical error) according to \cite{Shen11}, and hence the averaged Eddington ratio $\lambda_{\rm Edd}=L_{{\rm bol,(bright+faint)/2}}/L_{{\rm Edd}}$ $\sim$ 3.3$\times10^{-3} $, where $L_{\rm Edd}= 1.26\times 10^{38} M_{{\rm BH}}/M_{\odot}$ $\rm erg\, s^{-1}$. Here we emphasize that the \mgii\ line usually does not follow the breathing mode \citep{Shen13,Yang19}, i.e., the line width may not change with continuum variation, and whether there is an intrinsic $R-L$ is still unclear. Thus the black hole mass based on \mgii\ bears a larger uncertainty compared to \hb.

\subsection{Two turn-on \mgii\ CL candidates}
Figure \ref{fig:2MgII_CL} exhibits two tentative turn-on \mgii\ CL AGNs, and both of their bright states are selected as \mgii-emitters. Together with J1525+2920, we find that all three \mgii\ transitions occurred when $f_{\rm 3000\AA}$ is below $10^{-17}$ \ergcmsA (or $L_{\rm 3000\AA}< 10^{43.5}$ \erg), which is much fainter than normal \hb\ CL AGNs \citep{Macleod16,Yang18}. 

J0948+0050. The \mgii\ variability in this object is very significant with $\Delta f_{\mgii} = 224 \pm 18$ \ergs\ ($>$ 5$\sigma$). However, in the faint state, there is still a little remnant of the broad \mgii. The light curve shows a strong variability of $\sim$ 1 mag over a timescale of $\sim$ 10 yrs. 

J2244+0043. The \mgii\ variability in this object is relatively weak with $\Delta f_{\mgii} = 73 \pm 23$ \ergs\ ($>$ 3$\sigma$).  The $EW_{\mgii} = 18 \pm 7$ \AA\ is also small, which almost reaches the lower EW boundary of \mgii-emitters in Figure \ref{fig:selection}. The residuals of bright and faint epochs show an insignificant broad \hb\ line. The light curve indicates it varies $\sim$ 0.5 mag over a timescale of $\sim$ 10 yrs.

Due to the lack of obvious accompanied transitions (e.g., Iron and Helium lines) and verification from multi-epoch spectra, together with the concerns above, we classified them as tentative \mgii\ CL AGNs. If they keep further brightening, we would expect the appearance of broad \hb\ component.

\subsection{A CL sequence}\label{sec:cl_seq}
Recently, \cite{Guo19} demonstrated that the dramatic changes in broad \ha/\hb\ emission in the observationally-rare CL quasars are fully consistent with their photoionization model, and the theoretical CL sequence predicted by their model provides natural explanations for the persistence of broad \mgii\ in CL quasars defined on \ha/\hb\ and the rare population of broad \mgii-emitters (see their Figure 8).

\begin{figure*}[h]
\centering
\includegraphics[height=12.cm]{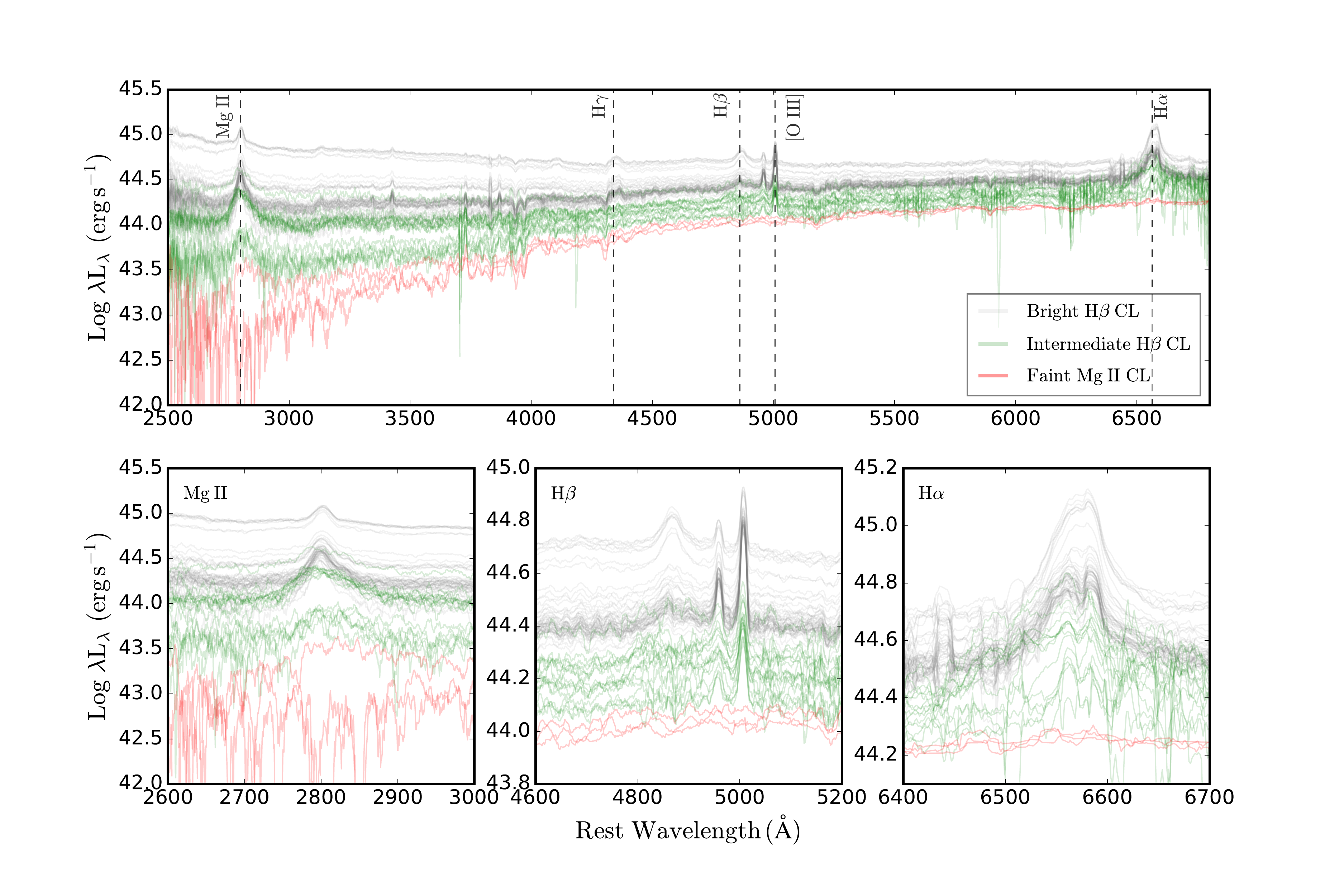}
\caption{The CL sequence. This compilation, presented sequentially with continuum luminosity decreasing, provides a representation of how the quasar type,
signalled by the varying strength of the different emission lines (i.e., \mgii\ and Balmer lines), may transition with luminosity. The sequence is constructed by three known CL objects from bright (SDSS J141324.27+530526.9, grey, z = 0.457 in \cite{Dexter19b}), intermediate (SDSS J022556.07+003026.7, green, z = 0.504, in \cite{Macleod16}), to faint states (J1525+2920, red, z = 0.449) at similar redshifts. All the multi-epoch spectra are obtained from SDSS database.  
}
\label{fig:evolution}
\end{figure*}

Here we recovered an observed CL sequence with real but three CL AGNs at different temporal evolution stages to confirm their prediction of the photoionizaiton model. In Figure \ref{fig:evolution}, two known multi-epoch \hb\ CL AGNs (J141324.27+530526.9, grey lines, z = 0.457 in \cite{Dexter19b} and J022556.07+003026.7, green lines, z = 0.504 in \cite{Macleod16}), together with our \mgii\ CL AGN (J1525+2920, red lines, z = 0.449), are selected to construct the observed CL sequence from bright, intermediate, to faint stages. The \mgii\ lines are sharing a similar profile for the faintest epoch of intermediate \hb\ CL AGN and the brightest epoch of the \mgii\ CL AGN, as well as the \mgii\ and Balmer lines in between the bright and intermediate CL AGNs. This allows temporal stages in three different CL AGNs, to link together to mimic the whole variability evolution in a CL object, compensating for the lack of multi-epoch spectroscopy across the full sequence in a single object. Although, the exact broad line width may be different or slightly affected by non-response effect in \mgii \citep{Guo19}, it would be trivial for demonstrating the concept of broad line disappearance. 

As shown in Figure \ref{fig:evolution}, when the broad Balmer lines almost disappear (e.g., become undetectable), the broad \mgii\ emission is still substantial. When the continuum luminosity continues to drop, broad \mgii\ eventually becomes too weak to be detectable. In this sequence, some faint epochs of the intermediate \hb\ CL AGN and brightest epoch of \mgii\ CL AGN are the so-called \mgii-emitters. Thus we speculate that the \mgii-emitter is more likely to be the transition quasar population where the quasar continuum and broad Balmer line flux had recently dropped by a large factor but the broad \mgii\ flux is still detectable on top of the stellar continuum. We also notice that the \mgii\ line always disappears later than \ha\ since \mgii\ has both less variability and suffers less contamination from the host galaxy. All these features are consistent with the theoretical CL sequence predicated by the photoionization model in \cite{Guo19}.

\section{Conclusion and Future work}\label{sec:con}

We have presented a systematic study of the spectroscopic variability of a sample of \mgii-emitters using multi-epoch spectra from the SDSS DR14. We have discovered the first unambiguous case of a \mgii\ CL AGN which is turning off in rest-frame 286 days, as well as two candidate turn-on \mgii\ CL AGNs. Together with two previously known \hb\ CL AGNs, we have constructed a unification sequence that represents different temporal stages of CL AGNs incorporating both broad Balmer-line and broad \mgii\ CL AGNs. We conclude that this CL AGN unification sequence is best explained by the photoionization model suggested by \cite{Guo19}, which indicates that most CL AGNs can be explained by the photoionization model. In this AGN CL sequence unification picture, \mgii\ emitters \citep{Roig14} are naturally explained as an intermediate stage of CL AGNs rather than a new AGN population.

We have also assembled a sample of 361 unique \mgii-emitters including 52 objects with multi-epoch spectra. They are useful for future searches of \mgii\ CL AGNs with dedicated spectroscopic time-domain surveys \citep[e.g., SDSS-V,][]{Kollmeier17,TheMSEScienceTeam2019}. With $\sim$ 6\% (3/52) \mgii\ CL AGNs in \mgii-emitters without accounting for selection incompleteness and selection biases, we would expect to discover about 20 \mgii\ CL AGNs assuming the full sample is monitored over a decade with an average cadence of $\sim$ a year. A significantly larger sample of CL AGNs will help put our results on firm statistical ground.

\begin{table*}
\centering
\begin{tabular}{lccccccccc}
\hline\hline
SDSS designation & Redshift &  $g$ band & Log $M_{\rm BH}$ & $\lambda_{\rm Edd}$ &Type& Plate &  MJD & Fiber &State \\
&&(mag)&($M_{\odot}$)&($10^{-3}$)&&&&&\\
(1)     & (2)  & (3)    & (4)    &   (5)    & (6)  & (7) & (8) &(9)& (10)\\
\hline
J152533.60$+$292012.1 & 0.449 &21.62$\pm$0.11& 8.0 $\pm$ 0.1 & 3.3 &turn-off&3879&55244&103&bright\\
$\dotfill$         		&           &&&&&3963&55659&731&faint\\
$\dotfill$         		&           &&&&&4721&55709&723&faint\\
\hline

J094810.92$+$005057.8& 0.624 &21.59$\pm$0.09& 7.4 $\pm$ 0.1 & 16.2 &turn-on&480&51989&99&faint\\
$\dotfill$         		&           &&&&&3827&55565&699&bright\\

J224448.72$+$004347.1& 0.637 &21.10$\pm$0.04& 7.8 $\pm$ 0.2 & 3.7 &turn-on&675&52590&489&faint\\
$\dotfill$         		&           &&&&&4204&55470&982&bright\\

\hline
\end{tabular}
\caption{Information for \mgii\ CL AGNs (hereafter J1525+2920, J0948+0050 and J2244+0043) from SDSS. Columns include the object name, redshift, $g-$band magnitude, \mgii-based black hole mass and 1 $\sigma$ statistical error, averaged Eddington ratio of bright and faint states, transition type, plate ID, MJD, fiber ID, and the luminosity state. Note that J0948+0050 was also collected in \mgii-emitter catalog of \cite{Roig14}. 
}
\label{tab:table1}
\end{table*}

\acknowledgments
We thank Y. Shen for helpful discussions. M.Y.S. acknowledges the support from NSFC-11603022. M.Z.K. is supported by Astronomical Union Foundation under grant No. U1831126 and Natural Science Foundation of Hebei Province No. A2019205100.

Funding for the Sloan Digital Sky Survey IV has been provided by the Alfred P. Sloan Foundation, the U.S. Department of Energy Office of Science, and the Participating Institutions. SDSS-IV acknowledges
support and resources from the Center for High-Performance Computing at
the University of Utah. The SDSS web site is www.sdss.org.

SDSS-IV is managed by the Astrophysical Research Consortium for the 
Participating Institutions of the SDSS Collaboration including the 
Brazilian Participation Group, the Carnegie Institution for Science, 
Carnegie Mellon University, the Chilean Participation Group, the French Participation Group, Harvard-Smithsonian Center for Astrophysics, 
Instituto de Astrof\'isica de Canarias, The Johns Hopkins University, 
Kavli Institute for the Physics and Mathematics of the Universe (IPMU) / 
University of Tokyo, Lawrence Berkeley National Laboratory, 
Leibniz Institut f\"ur Astrophysik Potsdam (AIP),  
Max-Planck-Institut f\"ur Astronomie (MPIA Heidelberg), 
Max-Planck-Institut f\"ur Astrophysik (MPA Garching), 
Max-Planck-Institut f\"ur Extraterrestrische Physik (MPE), 
National Astronomical Observatories of China, New Mexico State University, 
New York University, University of Notre Dame, 
Observat\'ario Nacional / MCTI, The Ohio State University, 
Pennsylvania State University, Shanghai Astronomical Observatory, 
United Kingdom Participation Group,
Universidad Nacional Aut\'onoma de M\'exico, University of Arizona, 
University of Colorado Boulder, University of Oxford, University of Portsmouth, 
University of Utah, University of Virginia, University of Washington, University of Wisconsin, 
Vanderbilt University, and Yale University.

The CSS survey is funded by the National Aeronautics and Space
Administration under Grant No. NNG05GF22G issued through the Science
Mission Directorate Near-Earth Objects Observations Program. 

\bibliography{ref}

%%%%%%%%%%%%%%%%%%% Tracking change %%%%%%%%%%%%%%%%%%%%

\end{CJK}
\end{document}